\documentstyle[multicol,aps,prl]{revtex}
\input{psfig.sty}
\def\eps{\varepsilon}

\begin{document}

\title{Anomalous exponents in the rapid-change model of the  \\
passive scalar advection in the order $\eps^{3}$.}

\author{L.\,Ts.\,Adzhemyan,  N.\,V.\,Antonov, V.\,A.\,Barinov,
Yu.\,S.\,Kabrits,  and A.\,N.\,Vasil'ev }

\address{Department of Theoretical Physics, St Petersburg University,
Uljanovskaja 1, St Petersburg---Petrodvorez, 198904, Russia}

\draft

\date{11 September 2000}

\maketitle

\begin{abstract}
Field theoretic renormalization group is applied to the Kraichnan model
of a passive scalar advected by the Gaussian velocity field with the
covariance
$\langle{\bf v}(t,{\bf x}){\bf v}(t',{\bf x})\rangle -
\langle{\bf v}(t,{\bf x}){\bf v}(t',{\bf x'})\rangle
\propto\delta(t-t')|{\bf x}-{\bf x'}|^{\eps}$.
Inertial-range anomalous exponents, related to the scaling dimensions of
tensor composite operators built of the scalar gradients, are calculated
to the order $\eps^{3}$ of the $\eps$ expansion. The nature and the
convergence of the $\eps$ expansion in the models of turbulence
are briefly discussed.
\end{abstract}
\pacs{PACS number(s): 47.27.$-$i, 47.10.+g, 05.10.Cc}

\begin{multicols}{2}

The investigation of intermittency and anomalous scaling in fully
developed turbulence remains essentially an open theoretical problem.
Both the natural and numerical experiments suggest that the deviation
from the predictions of the classical Kolmogorov theory
\cite{Legacy} is even more strongly pronounced for a passively
advected scalar field than for the velocity field itself; see, e.g.,
Ref.~\cite{Sree} and literature cited therein. At the same time,
the problem of passive advection appears to be easier tractable
theoretically: even simplified models describing the advection by
a ``synthetic'' velocity field with a given Gaussian statistics
reproduce many of the anomalous features of genuine turbulent
heat or mass transport observed in experiments. Therefore, the
problem of passive scalar advection, being of practical
importance in itself, may also be viewed as a starting point
in studying anomalous scaling in the turbulence on the whole.

Most progress has been achieved for the so-called rapid-change model
\cite{Kraich1}: for the first time, the anomalous exponents have been
calculated on the basis of a microscopic model and within regular
perturbation expansions; see
Refs.~\cite{Kraich1,Falk1,GK,Siggia,Pumir,%
VMF,RG,RG1,Juha,RG3,ReG,KJW,instanton} and references therein.

In that model, the advection of a passive scalar field $\theta(x)\equiv
\theta(t,{\bf x})$ is described by the stochastic equation
\begin{equation}
\partial _t\theta+ (v_{i}\partial_{i})  \theta
=\nu _0\Delta \theta+f,
\label{1}
\end{equation}
where $\partial _t \equiv \partial /\partial t$,
$\partial _i \equiv \partial /\partial x_{i}$, $\nu _0$
is the molecular diffusivity coefficient, $\Delta$ is the Laplace
operator, ${\bf v}(x)\equiv \{v_{i}(x)\}$ is the transverse (owing
to the incompressibility) velocity field, and $f\equiv f(x)$ is an
artificial Gaussian scalar noise with zero mean and correlator
\begin{equation}
\langle  f(x)  f(x')\rangle = \delta(t-t')\, C(r/L), \quad
r=|{\bf x}-{\bf x'}|.
\label{2}
\end{equation}
The parameter $L$ is an integral scale related
to the scalar noise, and $C(r/L)$ is some function finite as $L\to\infty$.
Without loss of generality, we take $L=\infty$ and $C(0)=1$.

In the real problem, the field  ${\bf v}(x)$ satisfies the Navier--Stokes
equation. In the rapid-change model it obeys a
Gaussian distribution with zero mean and correlator
\begin{eqnarray}
\langle v_{i}(x) v_{j}(x')\rangle = D_{0}\,
\frac{\delta(t-t')}{(2\pi)^d}
\int d{\bf k}\, P_{ij}({\bf k})\times
\nonumber \\
\times k^{-d-\eps}\, \exp [{\rm i}{\bf k}\cdot({\bf x}-{\bf x'})] ,
\label{3}
\end{eqnarray}
where $P_{ij}({\bf k}) = \delta _{ij} - k_i k_j / k^2$ is the
transverse projector, $k\equiv |{\bf k}|$, $D_{0}>0$ is an amplitude
factor, $d$ is the
dimensionality of the ${\bf x}$ space, and $0<\eps<2$ is a parameter
with the real (``Kolmogorov'') value $\eps=4/3$.
The infrared (IR) regularization is provided by
the cut-off in the integral (\ref{3}) from below at
$k\simeq m$, where $m\equiv 1/\ell$ is the reciprocal of another
integral scale $\ell$; the precise form of the cut-off is not essential.
The relation $D_{0}/\nu_0 =  \Lambda^{\eps} $ defines the
characteristic ultraviolet momentum scale $\Lambda$.

The issue of interest is, in particular, the behavior of the
equal-time structure functions
\begin{equation}
S_{n}(r) =\Big\langle[\theta(t,{\bf x})-\theta(t,{\bf x'})]^{n}\Big\rangle,
\quad  r =|{\bf x}-{\bf x'}|
\label{struc}
\end{equation}
in the inertial-convective range $\Lambda \gg 1/r \gg m$.

In the isotropic model (\ref{1})--(\ref{3}), the odd multipoint correlation
functions of the scalar field vanish, while the even equal-time
functions satisfy linear partial differential equations
\cite{Kraich1,Falk1,GK}.
The solution for the pair correlator is obtained
explicitly; it shows that the structure function $S_{2}$ is
finite for $m=0$ \cite{Kraich1}. The higher-order
correlators are not found explicitly, but their asymptotic
behavior for $m\to0$ can be extracted from the analysis of
the nontrivial zero modes of the corresponding differential
operators in the limits $1/d\to0$ \cite{Falk1},
$\eps\to0$ \cite{GK,Pumir}, or $\eps\to2$ \cite{Siggia,Pumir}.
It was shown that the structure functions in the
inertial-convective range exhibit anomalous scaling behavior:
\begin{equation}
S_{2n}(r)\propto D_{0}^{-n}\, r^{n(2-\eps)}\, (mr)^{\Delta_{2n}}
\label{HZ1}
\end{equation}
with negative anomalous exponents $\Delta_{n}$, whose first terms of the
expansion in $1/d$ \cite{Falk1} and $\eps$ \cite{GK} have the forms
\begin{eqnarray}
\Delta_{n}&=& -n(n-2)\eps/2d +O(1/d^{2}) =
\nonumber \\
&=& -n(n-2)\eps/2(d+2)+O(\eps^{2}).
\label{HZ3}
\end{eqnarray}

In paper \cite{RG}, the field theoretic renormalization
group (RG) and operator product expansion (OPE) were applied
to the model (\ref{1})--(\ref{3}). In the RG approach, the
anomalous scaling for the structure functions and
various pair correlators is established as a consequence of
the existence in the corresponding operator product expansions
of ``dangerous'' composite operators (powers of the local
dissipation rate), whose {\it negative} critical dimensions
determine the anomalous exponents $\Delta_{n}$.
The anomalous exponents were calculated in Ref.~\cite{RG}
in the order $\eps^{2}$ of the $\eps$ expansion for the arbitrary
value of $d$; generalization to the compressible case was given in
\cite{RG1,Juha}. The main advantage of the RG approach (apart from its
calculational efficiency) is the universality: it can equally be
applied to the case of finite correlation time \cite{RG3}.

In this paper, we present the anomalous exponents and other quantities
for the model (\ref{1})--(\ref{3}) in the order $\eps^{3}$.
Here we give only basic ideas and results; more exhaustive discussion
of the calculational technique will be given elsewhere.
A general review of the RG approach to the
statistical theory of turbulence can be found in Refs.~\cite{UFN,turbo};
the case of the Kraichnan model is discussed in \cite{RG} in detail.

The stochastic problem (\ref{1})--(\ref{3}) can be reformulated
as a multiplicatively renormalizable field theoretic model; the
corresponding RG equations have an IR attractive fixed point. This
implies existence of the infrared scaling behavior for all
correlation functions with certain scaling dimensions, calculated
as series in $\eps$ (in this sense, the exponent $\eps$ plays in the
RG approach the same role as the parameter $\eps=4-d$ does in the RG theory
of critical behavior). In particular, for the structure functions
(\ref{struc}), (\ref{HZ1}) in the IR asymptotic range ($\Lambda r \gg1$)
one obtains
\begin{equation}
S_{2n}(r)\propto D_{0}^{-n}\, r^{n(2-\eps)}\, \chi_{n}(mr).
\label{IRsca}
\end{equation}
The behavior of the scaling functions $\chi_{n}(mr)$ at $mr\to0$
(inertial-convective range) is obtained with the aid of the operator
product expansion:
\begin{equation}
\chi_{n}(mr)= \sum_{F} C_{F}\,  (mr)^{\Delta_{F}},
\label{OPE}
\end{equation}
where the sum runs over all possible composite operators $F$ entering
the OPE for a given structure function, $\Delta_{F}$ are their
critical dimensions, and $C_{F}$ are numerical coefficients analytical in
$(mr)^{2}$ and finite at $mr=0$.

The key role is played by the critical dimensions $\Delta_{nl}$,
associated with the tensor composite operators
\begin{equation}
F_{nl}= \partial_{i_{1}}\theta\cdots\partial_{i_{l}}\theta\,
(\partial_{i}\theta\partial_{i}\theta)^{p},
\label{Fnp}
\end{equation}
where $l$ is the number of the free vector indices and $n=l+2p$
is the total number of the fields $\theta$ entering the operator;
the vector indices of the symbol $F_{nl}$ are omitted.

The dimension $\Delta_{n}\equiv\Delta_{n0}$ of the scalar operator is
nothing other than the anomalous exponent in Eq. (\ref{HZ1});
see Ref.~\cite{RG}.
The dimensions with $l\ne0$ become relevant if the forcing (\ref{2})
is anisotropic: $\Delta_{nl}$ corresponds to the zero-mode
contribution to the $l$-th term of the Legendre decomposition
for the function $S_{n}$; see Ref.~\cite{RG3}.
They can be systematically calculated as series in $\eps$:
\begin{equation}
\Delta_{nl} = \sum_{k=1}^{\infty}  \, \Delta^{(k)}_{nl} \, \eps^{k},
\label{epsilon}
\end{equation}
with the first-order coefficient \cite{RG3}
\begin{equation}
\Delta^{(1)}_{nl} = -\frac{n\,(n-2)}{2(d+2)} +
\frac{(d+1)\,l\, (d+l-2)} {2(d-1)(d+2)}
\label{Qnp}
\end{equation}
(see also Refs.~\cite{ReG,KJW}; for $l=0$ this gives the result of \cite{GK},
while for $n=3$ and $l=1$ or 3 the result of \cite{Pumir} is recovered).

The coefficients $\Delta^{(2)}_{n0}$ and $\Delta^{(2)}_{n2}$ were obtained
in Ref.~\cite{RG} for any $n$ and $d$; the result for general $l$ is presented
in \cite{Juha}. In particular, one has
\begin{eqnarray}
\Delta^{(2)}_{nl}  &=&  n(n-2)(0.000203n -0.02976) -
\nonumber \\
&-& l^{2} (0.01732n + 0.01223)
\label{Qnp22}
\end{eqnarray}
for $d=2$ and
\begin{eqnarray}
\Delta^{(2)}_{nl}  &=& n(n-2)(0.00203n-0.00384 ) -
\nonumber \\
&-& l(l+1) (0.00710n - 0.00619)
\label{Qnp23}
\end{eqnarray}
for $d=3$ (analytical results are too cumbersome and will not be given here;
see Refs. \cite{RG,Juha}).

Now let us turn to the $O(\eps^{3})$ contribution. No analytical result
for it is available for general $d$; the numerical results have the forms
\end{multicols}
\begin{eqnarray}
\Delta^{(3)}_{nl} = n(n-2)(0.005472n^{2}+0.0649n+0.0647)
+l^{2} (-0.02161n^{2}-0.1023n+0.2406+0.01841 l^{2})
\label{Qnp32}
\end{eqnarray}
for  $d=2$ and
\begin{eqnarray}
\Delta^{(3)}_{nl} = n(n-2)(0.00140n^{2}+0.0199n+0.0343)
+l(l+1) \big(-0.00420n^{2}-0.0241n+0.0028(l^{2}+l+18)\big)
\label{Qnp33}
\end{eqnarray}
\begin{multicols}{2}
\noindent for $d=3$.
The quantity $\Delta^{(3)}_{nl}$ can be expanded as a series in $1/d$;
the coefficients of such expansion can be found, in principle, to any
given order. For general $n$ and $l$ to the order $1/d^{2}$ we have obtained
\end{multicols}
\begin{eqnarray}
\Delta_{nl} &=& \eps \big[ -n(n-2) (1-2/d) /2d + (l/2) (1-2/d+l/d
+2/d^{2}) \big] + 3\eps^{2} (n-2)(n-l) /4d^{2} +
\nonumber \\
&+& \eps^{3} (n-l) \big[ 1.74988(n-2)-0.624916l \big] /d^{2}.
\label{Qd}
\end{eqnarray}

\begin{multicols}{2}
Note that the $\eps^{2}$ and $\eps^{3}$ contributions decay for
$d\to\infty$ faster than $1/d$ in agreement with the $O(1/d)$ result
obtained in Ref. \cite{Falk1} for $\Delta_{n0}$. Moreover, from Eq.
(\ref{Qd}) it follows that the leading $O(1/d^{2})$ terms in these
contributions vanish for $n=l$, so that the decay at $d\to\infty$ becomes
even faster:
\end{multicols}
\begin{eqnarray}
\Delta_{nn} &=& \eps n/2 + n(n-1)\, \bigl\{ \eps\, /(d-1)(d+2) -
  \eps^{2} \big[ 1+ (2n-7)/d \big] / d^{3} -
\eps^{3} (3n-8) /2d^{4} \bigr\} +  O(\eps^{4}),
\label{Qdd}
\end{eqnarray}

\begin{multicols}{2}
\noindent with the accuracy of $O(1/d^{4})$.

We also recall that $\Delta_{20}=0$ to all orders in $\eps$ in agreement with
the exact solution for the second-order structure function \cite{Kraich1},
and that the exact nonperturbative result for $\Delta_{22}$ exists for all
$\eps$ and $d$ \cite{Falk1}.

For the isotropic model (\ref{1})--(\ref{3}), only scalar operators enter
the expansion (\ref{OPE}), the number of the fields $\theta$ in
the operators does not exceed the number of $\theta$'s on the left hand side,
and the leading term of the small-$mr$ behavior is given by the operator
with the minimal dimension $\Delta_{F}$. This allows one to identify the
anomalous exponent $\Delta_{n}$ in Eqs. (\ref{HZ1}) and (\ref{HZ3})
with the critical dimension $\Delta_{n0}$ of the scalar operator $F_{n0}$.

If the noise covariance (\ref{2}) involves some fixed constant vector
${\bf n}$ (large-scale anisotropy), the above results for the dimensions
$\Delta_{nl}$ do not change, but the operators with $l\ne0$ also enter
the right hand side of Eq. (\ref{OPE}) and give rise to contributions
proportional to $P_{l}(z)$, the $l$-th order Legendre polynomial,
$z$ being the angle between the vectors ${\bf n}$ and ${\bf r}$.
The odd structure functions $S_{2n+1}$
become nontrivial, and the leading term of their inertial-range behavior
is determined by the dimension $\Delta_{2n+1,1}$ of the vector operator
$F_{2n+1,1}$
\begin{equation}
S_{2n+1}(r)\propto D_{0}^{-n-1/2}\, r^{(n+1/2)(2-\eps)}\,
(mr)^{\Delta_{2n+1,1}}
\label{HZ13}
\end{equation}
(for more detail, see Refs. \cite{RG3}).

In Fig.~1, we show the dimension $\Delta_{4}$ (which determines the
anomalous exponent for $S_{4}$) for $d=3$ in the first, second, and
third orders in $\eps$. In Figs.~2 and 3, we show the ``anomaly''
$\gamma$, defined by the relation $S_{3}\propto r^{3-\gamma}$,
for $d=3$ and 2, respectively; note that the
$O(\eps^{2})$ curve lies above the $O(\eps)$ line for
$n=3$ and below it for $n=4$.
In the same Figures, we also present nonperturbative
results obtained for $n=4$ in Refs.~\cite{VMF} using numerical
simulations, and for $n=3$ in \cite{Pumir} using numerical
integration of the zero-mode equations
($\Delta_{4}=\zeta_{4}-2\zeta_{2}$ in the notation of \cite{VMF}
and $\gamma=3-\lambda$ in the notation of \cite{Pumir}).

An important issue which can be discussed on the example of the rapid-change
model is that of the nature and convergence properties of the $\eps$
expansions in models of turbulence and the possibility of their extrapolation
to finite values $\eps\sim1$. Figures 1--3 show that the agreement between
the $\eps$ expansion and nonperturbative results for small $\eps$ improves
when the higher-order terms are taken into account, but the deviation becomes
remarkable for $\eps\sim1$ and decreasing $d$. Furthermore, the
coefficients of the $\eps$ series appear more irregular for $d=2$
(see Fig. 3), while the forms of the nonperturbative results
\cite{Pumir,VMF} are not much affected by the choice of $d$.

Such behavior can be understood on the basis of the exact analytical
result for $\Delta_{22}$, which can be written in the form \cite{Falk1}
\begin{equation}
2 \Delta_{22} = -d-2+\eps + \sqrt{(\eps+\eps_{+})(\eps+\eps_{-})},
\label{koren}
\end{equation}
where
$$ \eps_{\pm} = (d^{2}+d+2 \pm \sqrt {8d(d+1)}\,)/(d-1). $$

It shows that the corresponding $\eps$ expansion has the finite radius of
convergence $\eps_{-}$, ranging from 0 to $\infty$ when $d$ varies from
1 to $\infty$; in particular, $\eps_{-}\simeq1.1$ for $d=2$ and
$\eps_{-}\simeq2.1$ for $d=3$. Hence the naive summation of the
$\eps$ expansion for $\Delta_{22}$ works only in the interval
$\eps<\eps_{-}$, which decreases almost linearly with $(d-1)$.
In order to recover the behavior of $\Delta_{22}$ from its $\eps$ series
for larger $\eps$, it is necessary to isolate explicitly
the singularity at $\eps_{-}$ in Eq. (\ref{koren}), thus changing
to a kind of improved $\eps$ expansion (whose radius of convergence
becomes $\eps_{+} \gg\eps_{-}$).
In practice, the first three terms of this improved expansion
approximate the exact result (\ref{koren}) equally good for
all $0<\eps<2$, both in two and three dimensions.

The difference with the models of critical phenomena, where $\eps$
series are always asymptotical, can be traced back to the fact that in
the rapid-change models, there is no factorial growth of the number of
diagrams in higher orders of the perturbation theory. The divergence
for $d\to1$ is naturally explained by the fact that the transverse vector
field does not exist in one dimension (we also recall that the RG
fixed point diverges at $d=1$; see Ref.~\cite{RG}). Thus it is natural to
assume that the series for higher-order exponents $\Delta_{nl}$ also
have finite radii of convergence with the behavior similar to that of
$\eps_{-}$. Therefore, in order to obtain reasonable predictions for
finite values of $\eps$, one should augment plain $\eps$ expansions
by the information about the location and character of the singularities.
Such information can be extracted from the asymptotical behavior of the
coefficients $\Delta^{(k)}_{nl}$ in Eq. (\ref{epsilon}) at large $k$.
To our knowledge, this problem has never been studied for dynamical models
like (\ref{1})--(\ref{3}); the instanton analysis developed in Refs.
\cite{instanton} has mostly been concentrated on the behavior of
the exponents in the limit $n\to\infty$. One can hope that the
implementation of the instanton calculus within the RG framework
will give the solution of this important problem.

N.V.A. acknowledges Juha~Honkonen,
Andrea~Mazzino and Paolo~Muratore~Ginanneschi for discussions
and the Center of Chaos and Turbulence Studies in the
Niels Bohr Institute for warm hospitality. The work was supported in part
by the Grant Center for Natural Sciences (Grant No. 97-0-14.1-30) and
Russian Foundation for Fundamental Research (Grant No. 99-02-16783).

\begin{figure}
\centerline{\psfig{file=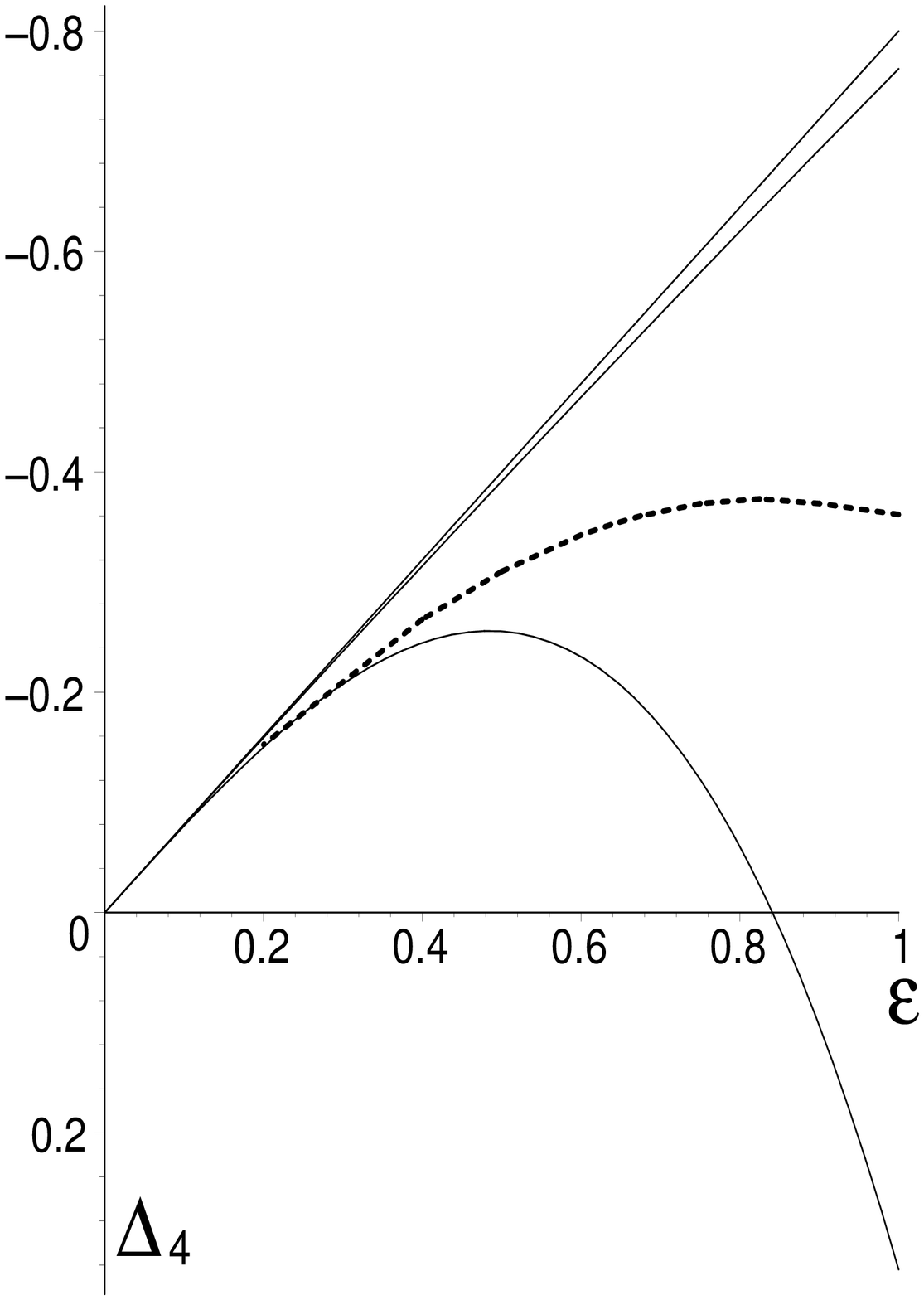,height=4cm,width=6cm}}
\caption{
The dimension $\Delta_{4}$ for $d=3$ {\em vs} $\eps$: the $O(\eps)$,
\mbox{\hspace{9.5cm}}
$O(\eps^{2})$ and $O(\eps^{3})$ approximations (from above to below).
\mbox{\hspace{9.5cm}}
Dashed line: numerical simulation by Refs. [8].} \label{fig1}
\end{figure}

\begin{figure}
\centerline{\psfig{file=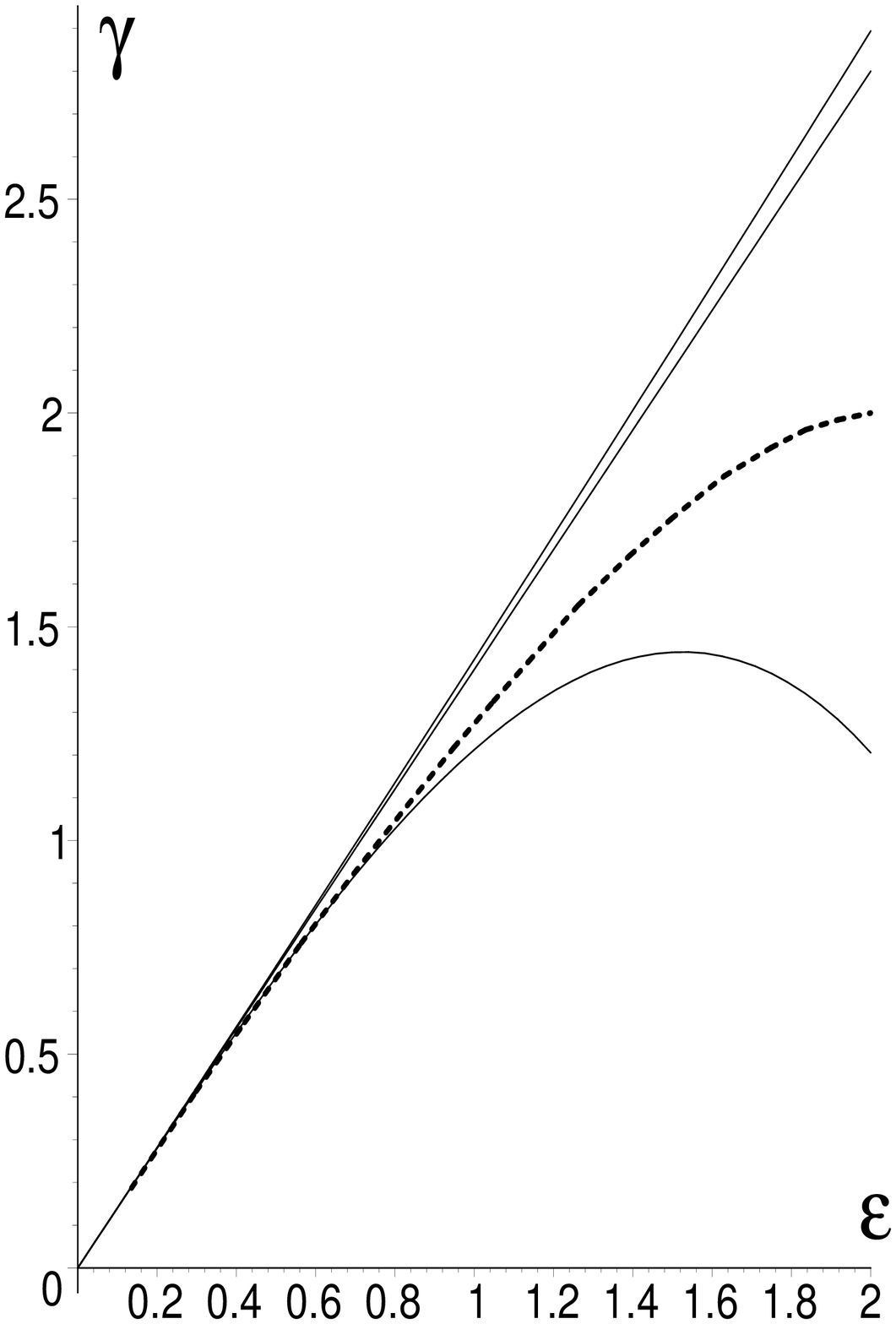,height=4cm,width=6cm}}
\caption{
The exponent $\gamma$ for $d=3$ {\em vs} $\eps$: the $O(\eps^{2})$,
\mbox{\hspace{9.5cm}}
$O(\eps)$ and $O(\eps^{3})$ approximations (from above to below).
\mbox{\hspace{9.5cm}}
Dashed line: numerical solution by Refs. [7].}
\label{fig2}
\end{figure}

\begin{figure}
\centerline{\psfig{file=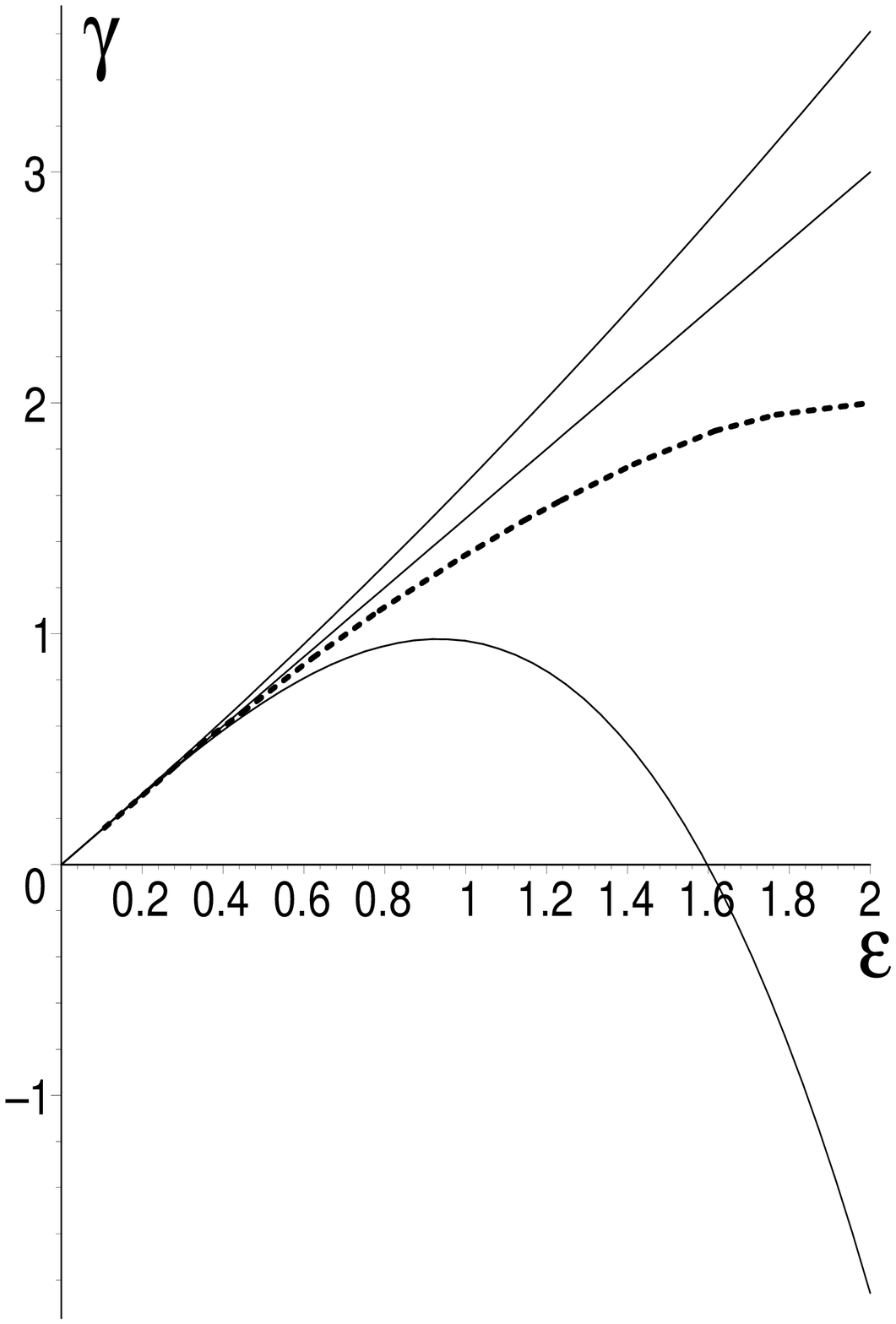,height=4cm,width=6cm}}
\caption{
The exponent $\gamma$ for $d=2$ {\em vs} $\eps$: the $O(\eps^{2})$,
\mbox{\hspace{9.5cm}}
$O(\eps)$ and $O(\eps^{3})$ approximations (from above to below).
\mbox{\hspace{9.5cm}}
Dashed line: numerical solution by Refs. [7].}
\label{fig3}
\end{figure}
\end{multicols}
\end{document}